\documentclass[aps, prd, twocolumn, lengthcheck, superscriptaddress, 
nofootinbib]{revtex4-1}

\usepackage{epsfig}
\usepackage[usenames]{color}
\usepackage{graphicx}
\usepackage{amsmath}
\usepackage{epstopdf}

\def\bi{\bibitem}

\def\la{\langle}\def\ra{\rangle}
\def\be{\begin{eqnarray}}\def\ee{\end{eqnarray}}
\def\lsim{\mathrel{\rlap{\lower3pt\hbox{\hskip1pt$\sim$}}
     \raise1pt\hbox{$<$}}} 
\def\gsim{\mathrel{\rlap{\lower3pt\hbox{\hskip1pt$\sim$}}
     \raise1pt\hbox{$>$}}} 
\def\del{\partial}

\allowdisplaybreaks

\begin{document}

\title{Dense Baryonic Matter Predicted in ``Pseudo-Conformal Model"}


\author{Mannque Rho}
\email{mannque.rho@ipht.fr}
\affiliation{Universit\'e Paris-Saclay, CNRS, CEA, Institut de Physique Th\'eorique, 91191, Gif-sur-Yvette, France }

\date{\today}

\begin{abstract}

The World-Class University/Hanyang Project launched in Korea in 2008 led to what's now called ``pseudo-conformal model" that addresses  dense compact-star matter  and is confronted in this short note with the presently available astrophysical observables, with focus on those from gravity waves. The predictions made nearly free of parameters by the model involving ``topology change" remain more or less intact ``un-torpedoed" by the data. 
 
\end{abstract}

\maketitle

\section{Introduction}
In 2008 the Korean Government launched a five-year ``World-Class University  (WCU)" Project, and the Hanyang University in Seoul was chosen as one of the projects to be under the directorship  of  Hyun Kyu Lee in the Physics Department. The objective of the WCU/Hanyang was to elevate the university in basic science to the world-class level, in anticipation of  the forth-coming establishment of an ambitious research institute called Institute of Basic Science (IBS).  The subject matter picked was ``Baryonic Matter under Extreme Conditions in the Universe " to be focused on superdense matter expected to be found in massive compact stars on the verge of  gravitational collapse. This subject matter was already one of the major themes in the Korea Institute of Advanced Studies (KIAS) in late 1990's and early 2000's while I was an invited professor in its School of Physics,  working in collaboration with  Hyun Kyu Lee, Dong-Pil Min and Byung-Yoon Park of Korea and Vicente Vento of  Spain, all at  the KIAS as visiting scholars.  

The property of dense baryonic matter in compact stars is in the realm of QCD involving both low and high densities. However QCD cannot access the density regimes, famously non-perturbative, of nuclear and compact star matter. Therefore there was no reliable theoretical tool to access the regimes concerned.  Neither could it be accessed experimentally since no accelerators probing dense matter at low temperature involved were available then. What initiated at the WCU/Hanyang Project was to build a {\it single} unified theoretical framework to explore {these} uncharted density regimes starting with what was explored in KIAS. The objective was to formulate an effective field theory approach with a minimal number of unknown parameters,   {\it post-dict correctly} the known nuclear matter properties at $n\sim  n_0\simeq 0.16$ fm$^{-3}$  and {\it predict} the terrestrial nuclear and compact-star properties that were yet to be measured. It was, in our mind, in anticipation of what is to be studied at the costly RIB machine ``RAON" approved to be constructed at the IBS. 

The status of the model in nuclear physics and astrophysics up to the early 2017  before the advent of the recent gravity-wave measurements was sketched in \cite{AAPPS}. The gravity-wave data  made feasible the direct confrontation of some of the predictions made then with the on-coming observables.

To the utter surprise -- and perhaps more to the incredulity -- of the workers in the field, what appears to be an over-simplified ``coarse-grained framework" with no parameter fiddling - in stark contrast to  the currently favored approaches of hybridizing with ``artificially revamped" quark descriptions -- has met, so far,  with no serious tension in explaining satisfactorily {\it all} up-to-date available data. In this paper, I will list the most relevant observables -- there are too many to be fair to all -- , both nuclear and astrophysical, just show how the predictions that followed from the formulation initiated at the KIAS and pursued at the WCU/Hanyang fare, and how the possible discrepancies, if any, between what's predicted and what's measured can be reconciled within the model. It should be  stressed that the spirit of this presentation is basically different from the current  activities in the field where various sophisticated statistical analyses in the theoretical inputs and experimental results are focused on. {All the results I will give are found essentially in the two papers \cite{PKLR,PKLMR}\footnote{Which constituted an important part of the PhD thesis of Won-Gi Paeng.} that are extensively reviewed in \cite{Rev1,Rev2}.\footnote{Some trivial numerical errors committed in \cite{PKLMR} that remained in \cite{Rev1,Rev2} will be corrected in the predicted results cited in what follows.} } Only if necessary will I refer to the specific articles for more precision or explanation. Otherwise I will avoid entering into details as much as possible.

\section{G$n$EFT}

In going from nuclear matter to dense compact-star matter, as commonly believed, there must be present a transition, either a phase change or a just continuous crossover, from the (low) density regime, say, $\sim 2 n_0$,  of hadrons to the (high) density regime, say, $\gsim 6 n_0$ of compact stars. This transition is commonly referred to as ``hadron-quark continuity (HQC)" presumed -- but not proven -- to be encoded in QCD.  The strategy that was adopted in the WCU/Hanyang  was that this HQC could be effectuated by a  change in topology from baryons in the baryonic matter to fractionally-charged objects in the compact-star matter, an idea anchored on what's referred to as ``Cheshire Cat Principle (CCP)". This idea followed from the notion that in QCD, a nucleon can be described as a topological object, say, a skyrmion and half-skyrmions at large $N_c$ and at high density when put on crystal lattice. An early review on this matter  can be found in \cite{park-vento}.

The key idea of how to implement the skyrmion-half-skyrmion transition -- referred in what follows to as ``topology change"\footnote{The topology change involved here could be different in character from what's taking place in condensed matter systems.} -- as a mechanism for the HQC was worked out first in early 2000 but appeared in the literature a decade later~ \cite{LPR}\footnote{The publication of this work was delayed so long due to the referees' objections to the novel ideas developed in the paper, arguing that they are mere ``conjectures."}. 

The topology change involved here is best described in terms of skyrmions put on crystal lattice, although it is well known that the skyrmion-half-skyrmion changeover actually makes no sense. This is because whereas the 1/2-skyrmion phase can be justified on crystal lattice at high density (and large $N_c$ limit), low-density matter cannot be in crystal, so the transition, whether  bona-fide phase transition or smooth cross-over, cannot be established with skyrmions on crystal lattice~\cite{adam}. This of course does not mean that the crossover in the skyrmion description in a more general setting does not exist. In fact it is this point that was resolved in \cite{LPR}: It involves hidden local symmetry (HLS) and hidden scale symmetry (HSS) entering into the baryonic structure. The details given in \cite{Rev1,Rev2} on how the hidden symmetries must figure appear to be somewhat complicated at first sight\footnote{I believe this accounts for the lack of attention paid to this development in nuclear and astrophysical communities.} but the basic structure is rather simple  as I will try to explain.  See \cite{cusp} for more details. 

\begin{figure}[h]\centering 
\includegraphics[scale=0.65,angle=0]{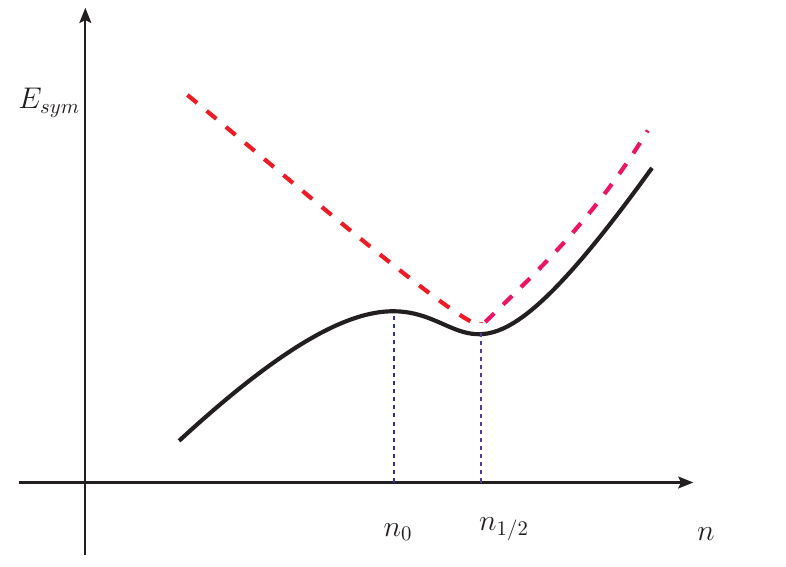}
\caption{Schematic illustration of the symmetry energy $E_{sym} (n)$  by the skyrmion crystal (dashed line) and by nucleon correlations dominated by the nuclear tensor forces (solid line). }
\label{LPR}
\end{figure}

The most crucial  ingredient for the topology change is the cusp structure in the symmetry energy (denoted $E_{sym}$). It reflects the isospin asymmetry in the energy functional $E(n)$. The cusp is seen when the nucleons are put on crystal lattice. It appears at the density, denoted $n_{1/2}$,  lying above the normal matter density $n_0$. Identified with as the putative HQC density, it is found to be in the range. 
\be
n_{\rm HQC}\sim n_{1/2}\approx  (2-4) n_0.
\ee  
This cusp  is displayed by the dotted red curve in the schematic figure, Fig. \ref{LPR}.
While one can reasonably assume that $E_{sym} (n) $ for $n\gsim n_{1/2}$ makes sense on the crystal lattice, the behavior for $n < n_{1/2}$ however cannot be taken seriously as mentioned above.  

First how does the topology change take place?

It is triggered on the crystal lattice by the bilinear quark condensate $\la\bar{q}q\ra$, when averaged, going to zero whereas the pion decay constant remaining non-zero. So it does not involve chiral symmetry restoration. It implies that the non-vanishing order parameter may be the quartic quark condensate $\la\bar{q}\bar{q} qq\ra$. There are arguments in the literature that such symmetry structure is at odds with 't Hooft anomaly constraints. But it remains controversial whether such a no-go theorem holds in the present case. In fact there are some cases in condensed matter where this no-go theorem does not seem to hold~\cite{garaud-babaev}.

To exploit this cusp structure requires knowing how the  topology change can be modified in reality.  Now how can the topology change be incorporated in a realistic theory? 

As shown in \cite{LPR}, it is the hidden symmetries that bring in heavy degrees of freedom to an effective Lagrangian. It is now recognized that Weinberg's chiral effective field theory (chiEFT) with the nucleons and pions as the {\it only} relevant degrees of freedom -- that will be called in what follows -- SchiEFT with ``S" standing for ``standard" works well with a suitable cutoff  $\Lambda_{sEFT} \lsim m_\rho$ up to the density $n_0$ and slightly higher.  It is bound  to break down at higher densities, say, $\gsim 2 n_0$. This success can be considered as a case where Weinberg ``Folk Theorem" on EFT is ``proven." What was shown in \cite{LPR} is that the vector mesons $V=(\rho, \omega)$ and the  scalar meson $(\sigma_d)$ which is related to what's known as ``conformal compensator"\footnote{A word on notation: $\chi$ is not to be confused with chiral symmetry.}, with the former endowed with ``vector manifestation fixed point" (VMFP)\footnote{At which the mass $m_\rho\to \epsilon\approx 0$.}~\cite{HLS} and the latter with the ``genuine dilaton" with an IR fixed point\footnote{At which the dilaton mass tends to $m_{\sigma_{d}}\to 0 $.}, 
enable one to go across, in the $E_{sym}$ (more generally the EoS), ``smoothly" from below  to above $n_{1/2}$. What is in action is the interplay between the VMFP and the nuclear tensor force that leads to $m_V\to 0$ at high density ($\gsim 25 n_0$) and the ``genuine dilaton" with an IR fixed point at $n_{IR} \gsim 25 n_0$ at which scale symmetry is restored. The net effect of the interplays at $n_{1/2}$ is displayed by the black solid curve  in Fig. \ref{LPR}. It will be shown later that the cusp, smoothed to an inflection, will play an important role in confronting some of the important  gravity-wave data, such as the tidal deformability, sound velocity etc.

The effective theory G$n$EFT detailed in the reviews~\cite{Rev1,Rev2}  is formulated with the Lagrangian  ${\cal L}_{\psi \pi {\rm HLS}\chi}$ with the HLS mesons  and the genuine dilaton (GD) scalar included as the relevant degrees of freedom in addition to the $\psi, \pi$ that figure in SchiEFT.  The heavy degrees of freedom (HDFs for short) are to mediate the crossover from hadrons to quark/gluons. In our approach, it is here that the topology change enters as a mechanism for hadron-quark continuity. 

Given the Lagrangian ${\cal L}_{\psi \pi {\rm HLS}\chi}$, there can be several ways of setting up a G$n$EFT. What is required is the implementation of the HQC at a density $n_{\rm HQC} > n_0$. 

It turns out to be feasible  to set up a scale-HLS-invariant Lagrangian\footnote{HLS is gauge-equivalent to non-linear sigma model, so chiral symmetry is encoded therein~\cite{HLS}.} with a power expansion going beyond the chiral expansion employed in the standard chiral EFT by  taking into account the hidden symmetries including HLS.  The expansion has been worked out to NLO in scale-chiral expansion following \cite{HLS,GD}. Unfortunately there are much too many parameters even at NLO that it has remained un-explored. 

An alternative approach exploited in the WCU/Hanyang program was to use the ``double-decimation" strategy developed in \cite{DD} which is to apply (Wilsonian) renormalization group approach to the strongly-correlated fermions on the Fermi sphere. The first decimation is made to Landau(-Migdal) Fermi liquid fixed point (FLFP) with the cutoff $\Lambda_{\rm FL}$ on top of the Fermi sea along the line developed in \cite{shankar} for electrons and then do the second decimation going beyond the FLFP. It was shown a long time ago that a chiral Lagrangian of the ${\cal L}_{\psi \pi {\rm HLS}\chi}$-type, somewhat simplified, can be mapped to Landau(-Migdal) Fermi liquid structure, which worked remarkably well at the FLFP level~\cite{FR,Song}.  This structure is incorporated into the G$n$EFT with the possibility of going beyond the FLFP
in the $V_{lowK}$-RG approach as developed by Tom Kuo with Gerry Brown and collaborators at Stony Brook. Tom Kuo's role in the initial development resulted in the crucial publication of  \cite{PKLR,PKLMR}. In the predictions discussed below, it will be primarily at the level of the FLFP approximation. The corrections in the $V_{lowK}$-RG will be quoted to justify the FLFP approximation. 

For those who are not familiar with the G$n$EFT strategy sketched above, let me just mention that this approach can be considered as a ``refined" version of covariant density functional approaches anchored on Hohenberg-Kohn theorem on DFT. The refinement, among others,  has to do with the replacement of the high dimension-field operators, injected (arbitrarily) to improve the Walecka-type linear model (e.g., the too high nuclear matter compression modulus { $K_0$}),  by the parameters of the Lagrangian with the dilaton condensate $\la\chi\ra$ encoded by the scale-chiral symmetry. The approach is free of arbitrariness and thermodynamically consistent~\cite{Song}.

\section{Predictions}

Here I will give the predictions obtained in \cite{PKLMR} and listed in \cite{Rev1,Rev2}. What's given involves no fiddling in the parameters in the Lagrangian ${\cal L}_{\psi \pi {\rm HLS}\sigma_d}$.  Only some numerical errors committed in  \cite{PKLMR} will be  corrected.

\subsection{Density regime $n \lsim n_0$}

First up to $n_{1/2}$ at which the HQC intervenes, what  is more or less equivalent  to what's given in  SchiEFT is reproduced by the mean-field of ${\cal L}_{\psi \pi {\rm HLS}\chi}$,  the parameters of which are controlled by BR scaling $\Phi$  sliding in density in the dilaton condensate $\la\chi\ra^\ast$ (where $\ast$ stands for the density dependence) known up to $n_0$.  At the equilibrium density $n_0$, one post-dicts\footnote{Just to give an idea what the significance of this result is, let me  quote what the present state-of-the art high-order (N$^{\gsim 2}$LO) SchiEFT calculation gets: $n_0 = 0.164 \pm 0.07$ and $E/A = -15.86 \pm 0.37 \pm 0.2 {\rm MeV}$. }
\be
n_0 = 0.16 {\rm fm}^{-3},\ E/A = -16.7\ {\rm MeV},\ K_0 = 250\ {\rm MeV}.\nonumber
\ee 
They  (and also the symmetry energy $J=E_{sym} (n_0)$ given in (\ref{L}) below)  are essentially the same as what's calculated in SchiEFT at N$^{\geq 2}$LO.
Here the only parameter needed is the mass of the ``genuine dilaton"  identified with $f_0(500)$. The  BR scaling relates the scaling of the dilaton condensate to that of the pion condensate
\be
\Phi (n)=f_{\sigma_d}^\ast/f_{\sigma_d}\simeq f_\pi^\ast/f_\pi
\ee
which is measured in deeply bound atomic nuclei $\Phi (n_0)\approx 0.8$. 

Roughly speaking the linear HLS with the BR scaling does what covariant density functional models with higher dimension operators do. The power of this approach is that it has thermodynamic consistency in addition to hidden local symmetry. It also captures higher chiral power terms, say, N$^3$LO in SchiEFT.

On the other hand, the symmetry energy slope $L$ could be different from what one gets in SchiEFT. This is because  of the onset of the cusp as shown in Fig.~\ref{LPR} at $n > n_{1/2}$. The cusp  as discussed in \cite{LPR} involves the tensor force structure controlled by the behavior of the HLS gauge coupling $g_\rho$ running in the RG flow toward  the vector manifestation $g_\rho\to 0$. If $n_{1/2}$ were not too far above $n_0$, then the slope of $E_{sym} (n)$ at $n_0$  would be inevitably  affected by the hidden cusp structure. I won't go for higher derivatives of $E_{sym}$ -- { such as $K_{\rm sym}$ with two derivatives} -- since they will depend more sensitively on where $n_{12}$ lies.

The G$n$EFT predicts for $n_{1/2}\sim (2-3)n_0$
\be
J \equiv E_{sym} (n_0)=30.2\ {\rm Mev}, \ L=67.8 \ {\rm MeV}\label{L}
\ee
to be compared with the SchiEFT results
\be
J=32.0\pm 1.1 \ {\rm Mev}, \ L=51.9\pm 7.9  \ {\rm MeV}. \label{schieft}
\ee
While $J$ is more or less the same as what SchiEFT gives,  ``soft" in the EoS,  $L$ is significantly greater than that of SchiEFT,  showing the (smooth) onset of hardness, tending toward what's observed in the PREX/Jefferson experiment $L=106\pm 37$~\cite{Jorge}.\footnote{I should make it clear that I will cite relevant references {\it only if} absolutely needed. There are a huge number of very well-written reviews of the current status following the development triggered by the gravity-wave data. I will not list them. For an up-to-date account, I will pick \cite{Jorge} for useful comments, whenever feasible, on the data.}  It is important to note that the behavior of $E_{sym} (n)$ near $n_0$ in G$n$EFT manifesting the ``pseudo-gap" behavior of the chiral condensate in the topology change  {\it predicts naturally} the soft-to-hard crossover tendency of the EoS at $\sim n_{1/2}$, what's attributed to the putative HQC in QCD.

\subsection{Density regime $n > n_{1/2}$}

Although the slope $L$ given in (\ref{L})  can be considered as a prediction, not as a pos-diction, of the PCM,  one cannot however have a great confidence in its precision. The reason is that it is the most difficult density regime in the EoS to theoretically control. At  $n_{1/2}$,  EFT valid at low density and perturbative QCD valid at high density ``meet."  Therefore the slope $L$ will be sensitive to the location with interplay of different degrees of freedom that can be treated with the least of confidence. This aspect will appear significantly in the tidal deformability $\Lambda$ measured at 1.4 $M_\odot$ and also in the sound velocity of the star.

While the $n\lsim n_{1/2}$ region is controlled essentially by the scaling factor $\Phi$, accessible both by theory and experiment, the topology change brings in major modifications in the properties of the Lagrangian ${\cal L}_{\psi \pi {\rm HLS}\chi}$. This is explained in terms of a series of ``Propositions" in \cite{Rev1}. I admit that some of them are superfluous or redundant and could be largely weeded out. 

Basically what happens is rather simple.  

Phenomenology in nuclear processes suggests the crossover density regime overlaps with the point $n_{DD}$ at which the double decimation is to be made~\cite{DD}. It has been taken to be~\cite{PKLR,PKLMR}
\be
n_{DD}\simeq n_{1/2}.
\ee
The primary mechanism that produces the cusp in the symmetry energy $E_{sym}$, namely, the skyrmion-1/2-skyrmion transition density, is driven in G$n$EFT by the nuclear tensor forces sliding with density, going to $\sim 0$ at the range most effective, say, $\sim 1$ fm  in nuclear interactions. What was required was that the VM fixed point density $n_{VM}$ be $n_{VM} \gsim 25 n_0$~\cite{PKLMR}, much greater than $\sim (6-7) n_0$  thought to be present in the core of massive stars.  This feature required that while the pion decay constant $f_\pi$  not go to zero at $n_{VM}$, it is the gauge coupling $g_\rho$ that should tend to zero~\cite{HLS}.\footnote{This feature which presumably takes place in temperature was not taken into account in heavy-ion experiments looking for the dropping $\rho$ mass near the chiral restoration temperature $T_c$. It led to the erroneous ``ruling out of BR scaling" following the NA60 data.} The scenario with $n_{VM}\gsim 25 n_0$ differs from $n_{VM}\sim 6 n_0$~\cite{ PKLR} in the prediction for the sound speed $v_s$ in compact stars. How the VM density $n_{VM}$ intervenes in the pseudo-conformal behavior of the sound velocity remains mysterious. 
 
 Another important property in $n\gsim n_{1/2}$ is that the { dilaton decay constant\footnote{From here on, I will use the linear conformal compensator field $\chi$ instead of the nonlinear field $\sigma_d$  for the dilaton field, $\chi=f_\chi e^{\sigma_d/f_\chi}$.} $f^\ast_\chi$ }  gets locked to the pion decay constant $f^\ast_\pi$ in the GD scheme~\cite{GD} and remains more or less constant
\be
{\rm\bf I}: f^\ast_\chi\simeq f^\ast_\pi\propto m_0 \ {\rm for} \ n > n_{1/2}\label{A}
\ee
where $m_0$ is a chiral symmetric mass of the quasiparticle in the 1/2-skyrmiom phase. This follows from the {\it emergent} parity doubling in the baryon structure\footnote{This feature differs from other parity-doubling scenarios where the symmetry is present intrinsically, not emergent, in the effective Lagrangian~\cite{PD}. It is not clear at the moment how this difference impacts on the properties of compact stars.}. One of the crucial consequences of this parity doubling is that the $U(2)$ symmetry for the $\rho$ and $\omega$, fairly good in $n < n_{1/2}$,  gets broken by the dynamics involved in the quasiparticle interactions with $\omega$ and $\chi$ exchanges in the 1/2-skyrmion phase
\be
m^\ast_\rho/m_\rho\neq m^\ast_\omega/m_\omega 
\ee
and leads to weakly interacting quasiparticles of bound 1/2-skyrmions with the mass\footnote{Since this is not generally known to nuclear theorists, let me point out that it was a particle theorist who showed that the two 1/2-skyrmions are bound to a single skyrmion -- a nucleon -- by the (hidden) monopoles~\cite{Cho}. I will speculate below how the suppression of the monopoles could lead to fractionized skyrmions mimicking quarks. } 
\be
{\rm\bf II}: m_Q^\ast \to f_\chi^\ast\to m_0.\label{B}
\ee
What's given in (\ref{A}) can also be obtained in what's referred to as ``dilaton-limit fixed point"~\cite{DLFP}  when ${\rm Tr} (\Sigma\bar{\Sigma})\to 0$ where $\Sigma=\frac{f_\pi}{f_\chi} e^{i\pi/f_\pi}\chi$ in the mean field of G$n$EFT. In that limit one finds
\be
{\rm\bf III}: g_A^\ast\to 1, \ f_\chi^\ast\to f_\pi^\ast.\label{C}
\ee

Since QCD cannot be solved nonperturbatively for the various limiting conditions,  the locations of  the DLFP, the vector manifestation (VM) fixed point,  the IR fixed point etc., though not too far apart,  are not precisely known.  For the issue concerned, i.e., the physics of compact stars, whether or not and where they overlap cannot  be addressed. They may, however, be irrelevant for the qualitative properties  we are interested in near the density regime of HQC. 

To be more quantitative,  one needs to go beyond the mean-field level approximation of G$n$EFT.  To do this, the $1/\bar{N}$ corrections to the Landau Fermi-liquid fixed point approximation -- in $V_{lowK}$ RG in the  double-decimation strategy~\cite{DD} -- could be made as described in \cite{PKLMR}. In this reference, a rather involved scaling behavior of the $\rho$ gauge coupling constant $g_\rho^\ast$ in the vicinity of the crossover density $n_{1/2}$ was used. Although it has not been checked in detail,  it seems most likely that such a complicated scaling behavior is unnecessary because it simply reflects how the gauge coupling moves toward the vector manifestation density $n_{VM}$ that lies way above the density involved in the star. This is indeed supported in the ``pseudo-conformal model (PCM)" (defined below)  used for making predictions.  

In listing the predictions made  in G$n$EFT there are two additional remarks to make:

First, the predictions have been made for the range of the crossover density 
\be
{\rm\bf IV}: 2 < n_{1/2}/n_0 < 4.\label{D}
\ee
The predictions are roughly the same within that range so I won't favor any specific values in between. The extremes $n_{1/2}/n_0=2$ and $4$ are somewhat disfavored although cannot be dismissed as we will see. 
Second the prediction made in the PCM  is checked with the double decimation $V_{lowK}$ RG only for $n_{1/2}/n_0 \sim 2$. It was concluded that the same  should hold for the range (\ref{D}). 

\begin{figure}[ht!]
\begin{center}
\includegraphics[width=8cm]{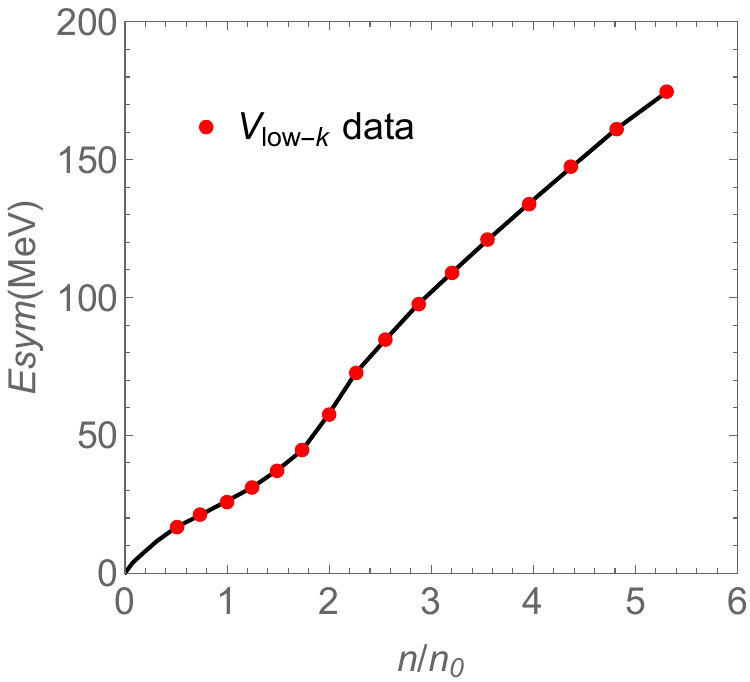}
\vskip -0.5cm
\caption{$E_{sym}$ (solid circle) obtained in the {\it full} $V_{lowk}$RG approach for $n_{1/2}= 2 n_0$.  It is reproduced  exactly by the pseudo-conformal model (solid line). Idem for $n_{1/2}\sim (2-4)n_0.$}\label{Esym}
\end{center}
\end{figure}

Second, the PCM\footnote{I must admit that the term ``pseudo-conformal" could be a misnomer. It simply indicates that conformal symmetry, both explicitly and spontaneously broken, {\it emerges} in dense matter driven by nuclear interactions.}  was constructed by replacing the $V_{lowK}$ RG for $n\geq  n_{1/2}$ in the energy density of the nucleon by two-parameter analytic form matched at $n=n_{1/2}$ to the $V_{lowK}$ for $n \leq n_{1/2}$. The matched energy density (PCM) is found to precisely reproduce $V_{lowK}$RG data for the whole range of density.  For example, in Fig.~\ref{Esym}, the symmetry energy $E_{sym}$ in the PCM (solid line) is shown to match exactly the full $V_{lowK}$RG. It also shows the higher-order terms beyond the mean-field approximation do indeed smoothen the cusp singularity -- schematically indicated in Fig.~\ref{LPR} -- as well as  correctly treat the density regime $ > n_{1/2}$. 
This result strongly suggests that the complicated scaling for the HLS gauge coupling used in the $V_{lowK}$ RG calculation  could well be made much simpler as mentioned above. 

As mentioned, the smooth matching of the skyrmion-1/2-skyrmion property at $n_{1/2}$ could be deceptive given the oversimplified joining of hadron-quark degrees of freedom.  The PCM may therefore hide complex structure lying just above the crossover density, say, in the density regime $\sim (2-4) n_0$. I will point this out in connection with some of the astrophysical observables to be discussed below.

What transpires from the properties ({\bf I})-({\bf IV}) incorporated into G$n$EFT for the PCM, is that the trace of the energy-momentum-tensor $\theta^\mu_\mu$  for $n\gsim n_{1/2}$ goes as
\be
\theta{_\mu^\mu} \propto \chi^4\to {\rm constant}.\label{theta}
\ee
This feature, the crucial element in the theory,  is reproduced in the $V_{lowK}RG$  double-decimation approach as shown in Fig.~\ref{Esym}.

\subsection{Predictions vs. observables}

Avoiding extensive references, both theory and experiment, I will list only those considered to be well determined to the extent that it is feasible with the source  from \cite{Jorge}. 
\begin{itemize}
\item {\bf Smoothed cusp of $E_{sym} (n)$ at $n\gsim n_{1/2}$:}

The bending-over of $E_{sym}$ influences the slope $L$ and induces the ``soft-to-stiff" changeover. It also plays a crucial role in giving rise to the pseudo-conformal sound velocity (to be addressed below). Although as stressed the detailed structure and magnitude cannot be precisely pinned down, its simplicity with intricate topology change in the jungle of theories (as depicted in \cite{cusp}) is a distinctive prediction of the PCM. It is at odds with the PREX-II/Jefferson data which give generally stiff EoS, although there are some caveats~\cite{Jorge}. There are up-to-date no trustful experimental data to quantitatively compare with.  
\item  {\bf Maximum mass star: $M^{\rm max}$}:
\be
{\rm PCM\ prediiction}:   &&M^{max} \approx 2.05 M_\odot, \nonumber\\
&& \ R_{2.0}\approx 12.8 \ {\rm km},\nonumber\\
&& (n_{central}\approx 5.1 n_0),\\
{\rm PSR J0740+6620}:&& M^{max} =2.08\pm 0.07 M_\odot,  \nonumber\\
&&  R_{2.0} =12.35\pm 0.75 \ {\rm km}, \nonumber\\
&& (n_{central}= ??),\label{I}\\
\ee
No empirical data is known to be  available at present for the central density $n_{central}$. The only information on this quantity {\it inferred} -- not extracted --  from PSR J0740+6620 is violently at odds with the PCM prediction. I will address this issue below.
\item {\bf 1.44  $M_\odot$ star}:
\be
&{\rm PCM\ prediction}&: R_{1.44}\approx 12.8\ {\rm km}\nonumber\\
&{\rm PSR J0030 + 0451}&:  R_{1.44} =12.45\pm 0.65 \ km.\label{II}
\ee
The stunning agreements between the PCM predictions  and the NICER and XMM-Newton measurements -- with the exception of the sound velocity to be addressed below -- could not be accidental. Not only the maximum star mass comes out the same but also the radii agree. Furthermore the difference $\Delta R=R_{2.0}-R_{1.4}\approx 0$ in agreement with the data. We will note later this support of the PCM by the NICER/XMM-Newton has even more surprising implication on scale-chiral symmetry in nuclear medium so far unsuspected.
\end{itemize}

Let me make some further comments here on the PCM results. 

What's given falls in the range of $n_{1/2}\sim (2.5 - 3.0)n_0$. The maximum mass comes out to be $\sim 2.4\ M_\odot$ for $n_{1/2}=4 n_0$. However at this crossover density, although other global properties are not drastically different from the lower values of $n_{1/2}$, the sound speed overshoots the causality bound with a more pronounced bump and the pressure greatly exceeds what's indicated in heavy-ion data. It seems to be ruled out in the PCM of G$n$EFT.

One observes that the radius comes out $\sim 12.8$ km in the wide range of the star mass and central density involved. Thus the stars of masses $\sim 1.4$ solar mass and $\sim 2.4$ solar mass have almost the same radius. This is in agreement with what's being observed in the gravity-wave data.  

\begin{itemize}
\item {\bf Tidal deformability $\Lambda_{1.4}$}

The  $\Lambda_{1.4}$ predicted in the PCM comes out to be  $\sim 550$, to be compared with 
$\Lambda_{1.4}=190^{+390}_{-120}$ (GW1700817). This may seem to signal a tension. However there is a basic difficulty in theoretically pinning down $\Lambda_{1.4}$. In the PCM, the density at which $\Lambda_{1.44}$ is measured is $\sim 2.4 n_0$. This density sits very close to where  the topology change takes place. It is here the SchiEFT is most likely to start breaking down as the cusp in $E_{sym}$ indicates and  the pQCD cannot access. This is an ``uncharted wilderness"  for theory. As can be seen in \cite{Rev1}, a small increase in the central density, say, from $2.3 n_0$ to $2.5 n_0$  (or increase in corresponding star mass), makes $\Lambda$ to drop to 420 while involving no change at all in radius. This means that the location of the HQC will strongly influence the $\Lambda$. {\it One can associate this behavior with the increase in attraction in going from $n_0$ toward $n_{1/2}$  in the cusp structure as one can see in the schematic plot Fig.~\ref{LPR}.}  This clearly suggests that it would be {\it extremely} difficult to theoretically pin down $\Lambda$ in the vicinity of the crossover regime.  

As noted below, the sound velocity has a complex ``bump" structure in the vicinity of the topology-change density. This is  due to the interplay, encoding the putative HQC,  between the hadronic degrees of freedom and the ``dual quark-gluon" degrees of freedom. This would complicate significantly the linking of $\Lambda_{1.4}$ to  the structure of the sound velocity below or near $n_{1/2}$. To give an example, let me quote \cite{jungle}  where the bump structure,  ``the slope, the hill, the drop, the swoosh, etc." -- associated with the possible phase structure of QCD is proposed to pin down  $\Lambda_{1.4}$ by up-coming measurements. The hope here is  to determine the possible phase transition near the HQC density. Given the theoretical wilderness inevitably involved, this seems a far-fetched endeavor.

In short, contrary to what's claimed by some workers in the field, ruling out an EoS based on the precise value of $\Lambda_{1.4}$ would be premature.
\item {\bf Sound speed $v_s$}
The most striking prediction of the PCM, so far not shared by other models, is the sound speed for $n\gsim n_{1/2}$. It {\it predicts} the pseudo-conformal sound speed
\be
v_s^{pcss}/c^2 \approx 1/3\ {\rm for}\ n \gsim n_{1/2}. 
\ee
It is not to be identified with the conformal sound speed $v_s^{conform}/c^2=1/3$ because the energy-momentum tensor is {\it not} traceless, i.e., scale symmetry is spontaneously broken. 

This prediction can be understood as follows.

As noted above, the quasiparticle mass $m^\ast_Q$ goes $\propto \la\chi\ra^\ast$ as the density goes above $n_{1/2}$ and the dilaton condensate  becomes independent of density, reaching $m_0$. This has to do with a delicate interplay between the attraction associated with the dilaton exchange and the $\omega$ repulsion which leads to the parity doubling. Where this interplay starts taking place cannot be pinned down precisely but  it must be in the density regime where the symmetry energy is involved, going from  $n_{1/2}$  to the core of massive stars, say, $\gsim 6n_0$. In this density regime, the Landau fixed-point approximation with ${\bar{N}}^{-1}=(\Lambda_F - k_f)/k_F \sim 1/k_F\to 0$ can be taken to be reliable.  One can then calculate the trace of the energy-momentum tensor  in the mean-field approximation of G$n$EFT, i.e., LFL fixed-point approximation,  which will become density-independent as given by (\ref{theta}). 
In this density range we will have
\be
\frac{\del}{\del n}\la\theta^\mu_\mu\ra=\frac{\del \epsilon(n)}{\del n}\Big(1-3\frac{v_s^2}{c^2}\Big) \approx 0
\ee
where $\epsilon (n)$ is the energy density and $v_s^2/c^2=\frac{\del P(n)}{\del n}/\frac{\del \epsilon (n)}{\del n}$. It is approximate since there can easily be terms that are compounded with EFT and pQCD at the point where the symmetry energy has the cusp structure. Since there is no Lee-Wick-type state, one must have
\be
\Big(1-3\frac{v_s^2}{c^2}\Big)\approx 0
\ee 
which gives the pseudo-conformal sound speed
\be
(v_s^{pcs}/c)^2 \approx 1/3.
\ee

The ``approximate zero" here stands for that it is pseudo-conformal with scale symmetry  broken both explicitly and spontaneously, the dilaton mass and the $\omega$ mass balancing so as to lead to parity-doubling in the dense system.
The true conformal velocity, within the model, should be reached only at a density much higher than that of the core density of the massive stars. Where precisely the conformality sets in is not relevant to the compact-star physics.

\begin{figure}[th]
\begin{center}
\includegraphics[width=8cm]{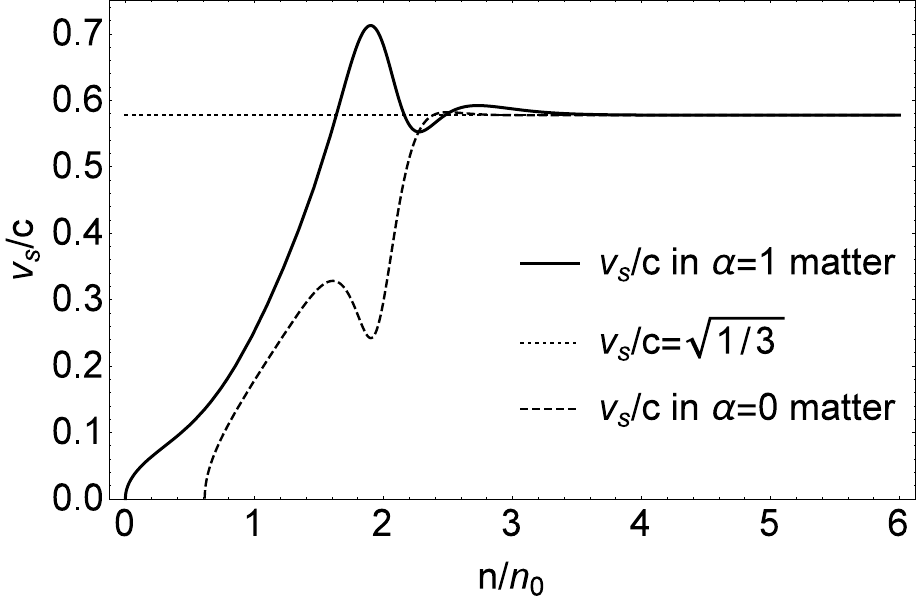}
\caption{
$v_s$ vs. density for $\alpha=0$ (nuclear matter) and $\alpha=1$ (neutron matter) in $V_{lowk}$ RG for $n_{1/2}=2 n_0$ and $v_{\rm vn}=25 n_0$.
 }\label{TEMT}
 \end{center}
\end{figure}

In Fig.~\ref{TEMT} is shown the sound speed  $v_s/c$ for $\alpha=0$ (nuclear matter) and 1(neutron matter) calculated in $V_{lowK}$RG for $n_{1/2}$. They are of the same form for $2< n_{1/2}/n_0 <4$ except for the slight shift in the density and the height of $v_s$. This result serves as an illustration of the arguments to follow.

What is noticeable is the large bump in $v_s$ in the vicinity of $n_{1/2}$ and the rapid convergence to the speed 1/3. The approximation involved on top of the pseudo-conformality would of course give fluctuations on top of  $v_{pcs}^2/c^2\approx 1/3$ but the point here is it is the pseudo-conformality that ``controls" the general structure. The large bump signals a complex interplay between hadronic  and non-hadronic degrees of freedom manifested through the pseudo-gap structure of  the chiral condensates. I will discuss below how the degrees of freedom in the core of the massive stars could masquerade ``deconfined quarks."

Though it's not directly connected with the star properties, a relevant and intriguing observation is what I would call ``quasibaryon" $g_A$ in nuclear matter. The effective $g_A$ in the Gamow-Teller transitions in nuclei, $g_A^{eff}$,  is observed to be $g_A^{eff}\approx 1$ from light nuclei to heavy nuclei and even to the dilaton-limit fixed point at $n\gsim 25 n_0$. It has been argued that an approximate scale invariance ``emerges" in nuclear interactions~\cite{gaeff}, in a way most likely related to the way $(v_s^{pcs}/c)^2\approx 1/3$ sets in precociously.

Now is there any indication in recent astrophysical observations for such a precocious onset of the pseudo-conformal sound velocity?

Up to date, there is no known ``smoking-gun" signal for the sound velocity from observations. In the literature, however, there are a gigantic number of articles  on the structure of sound velocity deduced from the gravity-wave data as well as theoretically. Some argue for phase transitions or continuous ones or simply no crossovers etc. Some extreme cases are discussed in \cite{cusp}. 
I won't go into this {\it wilderness} here.  Let me just describe one case which illustrates most transparently what can very well be involved.  

 Let's take  the case of NICER and XMM-Newton observables (NXN for short) discussed, namely (\ref{I}) and (\ref{II}).  This case brings out how puzzling the problem can be.

In \cite{bump-impact}, the properties of high density matter were inferred in most detailed analyses of the NXN data. Ruling out essentially all other scenarios, with or without phase changes, the authors arrive at the sound velocity (``H-bump") plotted in Fig.~\ref{bumps}. 
\begin{figure}[h]
    \centering
    \includegraphics[width=0.5\textwidth]{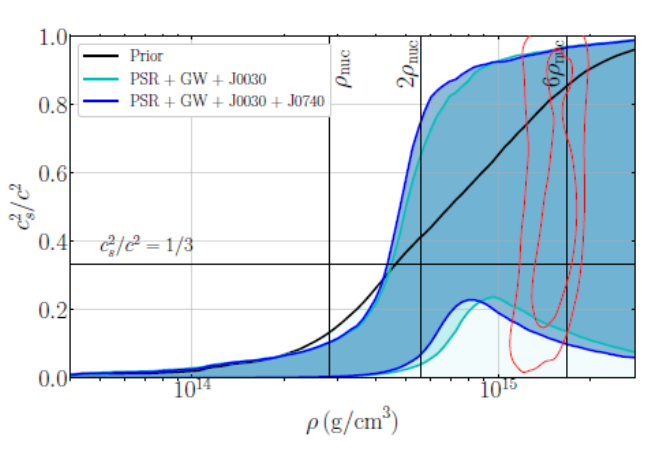}
    \caption{
  $v_s^2/c^2$ vs. $\rho$ in unit of $g/cm^3$ (the ``H-bump" scenario) taken from \cite{bump-impact}.}
\label{bumps}
\end{figure}
The central density and the maximum sound velocity inferred were
\be
n_{\rm cent}/n_0 &=& 3.0^{+1.6}_{-1.6},\nonumber\\
v_s^2/c^2  &=& 0.79^{+021}_{-0.20}.
\ee
While the star properties they took into account are exactly those reproduced by the PCM, i.e., (\ref{I}) and (\ref{II}),  the central density and the sound velocity are totally different from the PCM predictions.  One can understand the low central density accounting for the sound speed overshooting the conformal bound, characteristic of strongly interacting hadronic phase. In fact there are in the literature numerous scenarios anchored on a variety of density-functional approaches giving rise to the wilderness of one form or other in the sound velocity -- including bumps similar to the H-bump -- but I am not aware of any  that can survive the battery of bona-fide constraints coming from the current observations both in theory and experiment as  claimed by \cite{bump-impact}. 

A puzzle immediately raised is this:: How can the PCM with an emergent (pseudo-)conformality and the strong H-bump with no hint of conformal symmetry give the almost  {\it identical} global star properties (\ref{I}) and (\ref{II})?:  {\it The only statement one can make at this point is (A) either the sound velocity and the global star properties are totally unrelated or (B) there is something wrong either in the strong H-bump scenario or in the simple PCM structure?} The option (A) is hard to accept, so perhaps the option (B) is a plausible possibility. My bet is the option (B) and the H-bump scenario is at odds with nature.

In this connection, let me make a remark on the role of conformal symmetry in the sound velocity currently being discussed in the literature. This issue is a focused topic in the MDPI's Special Issue on ``Symmetries and Ultra Dense Matter of Compact Stars" being edited with contributions devoted to the issue. Without going into details let me just mention that there are a variety of models hybridizing hadronic degree of freedom and ``revamped" quark/gluon degrees of freedom at a density $n_{HQC}\gsim 2 n_0$. Some of the models such as quarkyonic  and holographic QCD do tend to see the  conformal symmetry (perhaps involving  percolation etc.) emerge  at certain density $\gsim n_{HQC}$ in going up in density~\cite{quarkyonic,sasaki,HQCD}.  Going down the density ladder from asymptotic density where $v_{conf}^2/c^2=1/3$,  one seems to observe the approximate conformality which persists down to the crossover regime where the big bump develops as it does in the PCM~\cite{pQCD}. This may represent a microscopic rendition of HQC in contrast to the PCM which presents a coarse-grained picture permeating in dense medium.  {This point is evidenced in Fig.~6 in \cite{Quark} where the results of quarkyonic models are compared with the PCM prediction Fig.~\ref{TEMT}.}
\end{itemize}

\section{Conclusion: Duck story}

{Briefly summarized, I have shown how to go from low density to high density capturing the {\it putative} hadron-quark continuity (HQC) by formulating baryonic matter as Landau-Migdal Fermi-liquid matter resulting  via renormalization group~\cite{shankar}. It is a sort of generalized density-functional approach (\`a la Hohenberg-Kohn theorem), implementing heavy degrees of freedom in terms of hidden symmetries involving a mass scale above that given by standard chiral EFT which is shown to be valid at nuclear matter density.  The resulting effective field theory, G$n$EFT,  exploits the possibility of simulating via duality the HQC in terms of a topology change from skyrmions at low density  $\sim n_0$ to 1/2-skyrmions at high density $\sim 6 n_0$. The resulting EoS has so far successfully accounted for nuclear matter as well as dense compact star matter. The structure that is arrived at in compact-star matter, coined as pseudo-conformality, can be considered as a coarse-grained description of  the hadrons-to-quarks changeover, e.g., quarkyonic {\it ``IdylliQ"}~\cite{larry}, captured in terms of ``emergent" scale symmetry permeating from low to high density.

The formulation made so far is valid at zero temperature. Upcoming terrestrial laboratory observations complimentary to astrophysical data, e.g, at FAIR of GSI, however, will necessarily involve relatively high temperature. It remains to be  formulated in the G$n$EFT framework to meet the conditions of the terrestrial laboratoris.  How topology enters in the hot {\it and} dense matter is a totally open issue as indicated in recent puzzling manifestations of scale invariance at high temperature~\cite{alexandru}. }

Finally I touch on fractionalized ``quasibaryon" structure inside the core of the massive star.

When a paper appeared in 2020~\cite{duck} with the suggestion that the cores of the most massive neutron stars are characterized by approximate conformal symmetry, with the speed of sound $v_s^2/c^2\to 1/3$, the polytropic index $\gamma= {\rm d\ ln} \ p/{\rm d\  ln}\ \epsilon\to 1$ and the normalized trace anomaly $\Delta=(\epsilon-3p)/(3\epsilon)\to \delta\approx 0$, indicating that the cores are most likely populated by deconfined fractionally charged objects, identified as quarks, those quantities were quickly calculated in the PCM formulated in 2017~\cite{PKLMR}. I considered this as a prediction of the PCM.  The predicted results~\cite{duckstory} were quite consistent with the conclusion of \cite{duck}.

Now the question was this: Given the degrees of freedom in the PCM are quas-ibaryons, albeit fractionaiized, how do they carry the characteristics of fractionally charged quarks? 

I do not have an  immediate answer to this question. But there are certain ideas that could lead to an understanding of this puzzle~\cite{fractionization}. One of them is this: In the skyrmion-half-skyrmion crystal simulation, the half-skyrmions ``confined" into a skyrmion by monopoles~\cite{Cho} could be liberated at high density and propagate freely with little interactions as seen in skyrmion crystals~\cite{PKLMR}. Two half skyrmions can then be rearranged into three 1/3-charged objects as in a schematic model~\cite{vento}. In fact in condensed matter physics, with domain walls, there can be stacks of sheets containing deconfined fractionally charged objects behaving like ``deconfined quarks" coming from the bulk in which the objects are confined~\cite{wall}.

This reminds one of the ``duck test": {\it If it looks like a duck, swims like a duck, and quacks like a duck, it probably is a duck."}

 \end{document}